\documentclass{jfm}
\usepackage[T1]{fontenc}
\usepackage[latin9]{inputenc}
\usepackage{xcolor}
\usepackage{verbatim}
\usepackage{units}
\usepackage{amsmath}
\usepackage{amssymb}
\usepackage{graphicx}
\usepackage{setspace}
\usepackage{esint}
\usepackage[authoryear]{natbib}
\PassOptionsToPackage{normalem}{ulem}
\usepackage{ulem}

\makeatletter

\providecommand{\tabularnewline}{\\}
\providecolor{lyxadded}{rgb}{0,0,1}
\providecolor{lyxdeleted}{rgb}{1,0,0}

\def\e{\mathrm{e}}
\def\i{\mathrm{i}}

\def\vec#1{\boldsymbol{\mathit{#1}}}
\def\underline#1{\boldsymbol{\mathsf{#1}}}

\affiliation{Applied Mathematics Research Centre,\\
Coventry University, Priory Street, Coventry CV1 5FB, UK}
\pubyear{2014}
\volume{760}
\pagerange{387--406}
\ifCUPmtlplainloaded \else
  \IfFileExists{amsbsy.sty}
    {\typeout{^^JFound the 'amsbsy' package on the system, using it.^^J}
}

\begin{document}

\title[2D nonlinear travelling waves in MHD channel flow]
{Two-dimensional nonlinear travelling waves \\
in magnetohydrodynamic channel flow}

\author[J. Hagan and J. Priede]
{Jonathan Hagan and J\=anis Priede} 

\date{11 December 2013; revised 12 September 2014; accepted 17 October 2014}

\maketitle

\begin{abstract}
This study is concerned with the stability of a flow of viscous conducting
liquid driven by a pressure gradient in the channel between two parallel
walls subject to a transverse magnetic field. Although the magnetic
field has a strong stabilizing effect, this flow, similarly to its
hydrodynamic counterpart -- plane Poiseuille flow -- is known to become
turbulent significantly below the threshold predicted by linear stability
theory. We investigate the effect of the magnetic field on two-dimensional
nonlinear travelling-wave states which are found at substantially
subcritical Reynolds numbers starting from $\textit{Re}{}_{n}=2939$
without the magnetic field and from $\textit{Re}{}_{n}\sim6.50\times10^{3}\textit{Ha}$
in a sufficiently strong magnetic field defined by the Hartmann number
$\textit{Ha}.$ Although the latter value is by a factor of seven
lower than the linear stability threshold $\textit{Re}{}_{l}\sim4.83\times10^{4}\textit{Ha},$
it is still more than an order of magnitude higher than the experimentally
observed value for the onset of turbulence in the magnetohydrodynamic
(MHD) channel flow. 
\end{abstract}

\section{Introduction}

The flow of viscous incompressible liquid driven by a constant pressure
gradient in the channel between two parallel walls, which is generally
known as plane Poiseuille or simply channel flow, is one of the simplest
and most extensively studied models of hydrodynamic instabilities
and transition to turbulence in shear flows \citep{Yaglom-12}. The
development of turbulence in the magnetohydrodynamic (MHD) counterpart
of this flow, which is known as Hartmann flow and arises when a conducting
liquid flows in the presence of a transverse magnetic field, is currently
not so well understood. The MHD channel flow, which was first described
theoretically by \citet{Hart37} and then studied experimentally by
\citet{Hart-Laz37} is still an active subject of research \citep{Hagan-Priede-13b,Krasnov-etal13}. 

Linear stability of Hartmann flow was first analysed by \citet{Lock55},
who showed that the magnetic field has a strong stabilizing effect
which results in the critical Reynolds number increasing from $\textit{Re}_{l}=5772.2$
without the magnetic field to $\textit{Re}_{l}\sim50\,000\textit{Ha}$
for Hartmann numbers $\textit{Ha}\gtrsim20.$ The critical Reynolds
number for the linear stability threshold based on the Hartmann layer
thickness $R_{l}=\textit{Re}_{l}/\textit{Ha}\approx50\,000$ is more
than two orders of magnitude higher than $R_{t}\approx225$ at which
transition from turbulent to laminar MHD channel flow was experimentally
observed by \citet{Mur53}. This extremely high critical Reynolds
number found by \citet{Lock55}, which is typical for exponential
velocity profiles \citep{Rob67,Drazin-Reid-81,Pot07}, has been confirmed
by a number of more accurate subsequent studies starting with \citet{Likhachev-76},
who found $R_{l}\approx48\,310,$ which is very close to the highly
accurate $R_{l}\approx48\,311.016$ obtained later by \citet{Takashima-96}.
Linear stability analysis of a single Hartmann layer has been carried
out by \citet{Ling-Albo-99} using a relatively rough numerical approach
which produced $R_{l}\approx48\,250.$ 

The result of \citet{Mur53} was, in turn, confirmed by \citet{BroLyk67}
as well as by \citet{Bra67}. A somewhat higher threshold value of
$R_{t}\approx380$ was found in the latest experiment on the transition
to turbulence in the Hartmann layer by \citet{MorAlb04}. This value
was supported by the accompanying numerical study of \citet{Krasnov-etal04},
who reported $R_{t}$ in the range between $350$ and $400$ (for
more detailed discussion of these results, see the recent review by
\citealt{Zikanov-etal14}).

A possible cause of the discrepancy between the theory and experiment
was suggested by \citet{Lock55} himself who conjectured that it may
be due to the finite amplitude of the disturbance which is not taken
into account by the linear stability analysis. This conjecture was
supported by the weakly nonlinear stability analysis of a physically
similar asymptotic suction boundary layer, which was found by \citet{Hocking-75}
and \citet{Likhachev-76} to be subcritically unstable with respect
to small but finite-amplitude disturbances. \citet{MorAlb03} later
confirmed the same to be the case also for the Hartmann layer also.
The first quantitative results concerning finite-amplitude subcritical
travelling waves in Hartmann flow were reported by \citet{LifSht80}.
Assuming the perturbation to be in the form of a single harmonic,
which is known as the mean-field approximation, they found such two-dimensional
(2D) travelling waves to exist down to the local Reynolds number $R_{n}=\textit{Re}_{n}/\textit{Ha}\approx12\,300.$
The use of this approximation is endorsed by its unexpectedly good
performance in the non-magnetic case, where it produces $\textit{Re}_{n}\approx2825$
\citep{Soibelman-Meiron-91}, which differs by less than $4\%$ from
the accurate result $\textit{Re}_{n}\approx2939$ \citep{Casas-Jorba-12}.
On the other hand, this approximation is known to be only qualitatively
correct even in the weakly nonlinear limit where it overestimates
the first Landau coefficient, which determines the evolution of small
finite-amplitude disturbances, by about $30\%$ \citep{Reynolds-Potter-67}.

Alternatively, there have been attempts to explain the transition
to turbulence in Hartmann flow  by the energy stability and the transient
growth theories. Although the former formally applies to arbitrary
disturbance amplitudes, it is essentially a linear and amplitude-independent
approach because the nonlinear term neither produces nor dissipates
the energy and, thus, drops out of the disturbance energy balance.
Using this approach \citet{Ling-Albo-99} found that the Hartmann
layer is energetically stable, \emph{i.e.,} all disturbances decay
at any time, when the local Reynolds number is below $R_{e}\approx26,$
which is almost  an order of magnitude lower than the experimentally
observed threshold. As demonstrated in the numerical study by \citet{Krasnov-etal04},
also the optimal transient growth which has been studied for the Hartmann
boundary layer by \citet{GerVar-02} and for the whole Hartmann flow
by \citet{AirCas04}. 

Transition to turbulence is essentially a nonlinear process which
is mediated by the equilibrium states that may exist besides the laminar
base flow at sufficiently high Reynolds numbers. However, though such
equilibrium states are usually unstable, they may have stable manifolds
that approach very close to the laminar base flow \citep{Chapman-02}.
Thus, a small finite-amplitude perturbation can easily bring the flow
into the stable manifold of such an equilibrium state. First, the
initial perturbation is amplified as it is attracted to the equilibrium
state along the stable manifold. Second, when the flow gets sufficiently
close to the equilibrium state, it is repelled along the unstable
manifold leading to another state. This may result in the flow wandering
between a number of such unstable equilibrium states which form the
so-called `skeleton of turbulence'. For the significance of such states
in turbulent flows, see the review by \citet{Kawahara-etal12}. It
has to be noted that the existence of multiple equilibrium states
is inherently a nonlinear effect which has little to do with the non-normality
linear problem underlying the transient growth \citep{Wal95}. Moreover,
as the evolution of the flow is determined by its initial state, it
is not the transient growth but rather the finite amplitude of the
initial perturbation which is required for the flow to get into the
basin of attraction of another equilibrium state. Even the so-called
optimal perturbations formed by the streamwise rolls, which lie at
the heart of the transient growth mechanism, require an additional
\emph{finite-amplitude} perturbation to reach turbulent attractor.
The breakdown of the streamwise rolls cannot be explained by their
apparent linear instability because because no infinitesimal disturbance
can reach finite amplitude required for the transition to turbulence
over the finite lifetime of these rolls. The gap between the linear
transient growth and the nonlinear dynamical system approach is bridged
by the optimisation approach recently reviewed by \citet{Kerswell-etal14}.

The present study, constituting the first step of a fully nonlinear
stability analysis, is concerned with finding 2D travelling-wave states
in Hartmann flow. Starting from plane Poiseuille flow, we trace such
subcritical equilibrium states by gradually increasing the magnetic
field. Using an accurate numerical method based on the Chebyshev collocation
approximation and a sufficiently large number of harmonics we find
that such states extend to the local subcritical Reynolds number $R_{n}\approx6500$
which is almost  a factor of two smaller than that predicted by the
mean-field approximation \citep{LifSht80}. 

The paper is organized as follows. The problem is formulated in $\S$\ref{sec:prob}.
In $\S$\ref{sec:theory} we present theoretical background concerning
2D nonlinear travelling-wave states. The Numerical method and the
solution procedure are outlined in $\S$\ref{sec:Num}. In $\S$\ref{sec:Results}
we present and discuss numerical results concerning 2D nonlinear travelling
waves and their linear stability with respect to 2D superharmonic
disturbances. The paper is concluded with a summary in $\S$\ref{sec:Sum}.

\section{\label{sec:prob}Formulation of problem}

\begin{figure}
\begin{centering}
\includegraphics[bb=80bp 100bp 380bp 280bp,clip,width=0.66\textwidth]{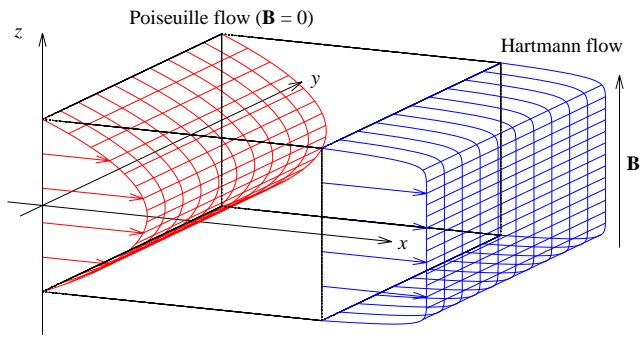} 
\par\end{centering}

\protect\caption{\label{fig:sketch}Sketch of the problem showing velocity profiles
of Poiseuille and Hartmann flows.}
\end{figure}

Consider the flow of an incompressible viscous electrically conducting
liquid with density $\rho,$ kinematic viscosity $\nu$ and electrical
conductivity $\sigma$ driven by a constant gradient of pressure $p$
in a channel of  width $2h$ between two parallel walls in the presence
of a transverse homogeneous magnetic field $\vec{B}$. The velocity
distribution of the flow $\vec{v}(\vec{r},t)$ is governed by the
Navier-Stokes equation
\begin{equation}
\partial_{t}\vec{v}+(\vec{v}\cdot\vec{\nabla})\vec{v}=-\rho^{-1}\vec{\nabla}p+\nu\vec{\nabla}^{2}\vec{v}+\rho^{-1}\vec{f},\label{eq:NS}
\end{equation}
where $\vec{f}=\vec{j}\times\vec{B}$ is the electromagnetic body
force containing the induced electric current $\vec{j},$ which is
governed by Ohm's law for a moving medium 
\begin{equation}
\vec{j}=\sigma(\vec{E}+\vec{v}\times\vec{B}),\label{eq:Ohm}
\end{equation}
where $\vec{E}$ is the electric field in the stationary frame of
reference. The flow is assumed to be sufficiently slow that the induced
magnetic field is negligible relative to the imposed one. This supposes
a small magnetic Reynolds number $\textit{Rm}=\mu_{0}\sigma v_{0}L\ll1,$
where $\mu_{0}$ is the permeability of vacuum and $v_{0}$ and $L$
are the characteristic velocity and length scale of the flow, respectively.
Using the thickness of the Hartmann layer $L=\delta=B^{-1}\sqrt{\rho\nu/\sigma}$
as the relevant length scale in strong magnetic field, we have $\textit{Rm}\sim R\textit{Pm},$
where $R=\textit{Re}/\textit{Ha}$ is a local Reynolds number based
on $\delta$ and $\textit{Pm}$ is the magnetic Prandtl number. Taking
into account that for typical liquid metals $\textit{Pm}\sim10^{-5},$
the constraint on $\textit{Rm}$ translates into $R\ll\textit{Pm}^{-1}\sim10^{5}.$
In addition, we assume that the characteristic time of velocity variation
$\tau\sim L/v_{0}$ is much longer than the magnetic diffusion time
$\tau_{m}=\mu_{0}\sigma L^{2}.$ This is also satisfied by the constraint
on $\mathbf{\textit{Rm}}$ and thus allows us to use the quasi-stationary
approximation leading to $\vec{E}=-\vec{\nabla}\phi,$ where $\phi$
is the electrostatic potential \citep{Rob67}. 

The velocity and current satisfy mass and charge conservation $\vec{\nabla}\cdot\vec{v}=\vec{\nabla}\cdot\vec{j}=0.$
Applying the latter to  Ohm's law (\ref{eq:Ohm}) yields 
\begin{equation}
\vec{\nabla}^{2}\phi=\vec{B}\cdot\vec{\omega},\label{eq:phi}
\end{equation}
where $\vec{\omega}=\vec{\nabla}\times\vec{v}$ is vorticity. At the
channel walls $S$, the normal $(n)$ and tangential $(\tau)$ velocity
components satisfy the impermeability and no-slip boundary conditions
$\left.v_{n}\right\vert _{s}=0$ and $\left.v_{\tau}\right\vert _{s}=0.$ 

We employ right-handed Cartesian coordinates with the origin set at
the mid-height of the channel, and the $x$- and the $z$-axes directed,
respectively, against the applied pressure gradient $\vec{\nabla}p_{0}=P\vec{e}_{x}$
and along the magnetic field $\vec{B}=B\vec{e}_{z}$ so that the channel
walls are located at $z=\pm h,$ as shown in figure \ref{fig:sketch},
and the velocity is defined as $\vec{v}=(u,v,w).$ Subsequently, all
variables are non-dimensionalized by using $h,$ $h^{2}/\nu$ and
$Bh\nu$ as the length, time and electric potential scales, respectively.
The velocity is scaled by the viscous diffusion speed $\nu/h,$ which
we employ as the characteristic velocity instead of the commonly used
centreline velocity.

The problem admits a rectilinear base flow 
\begin{equation}
\vec{v}_{0}(z)=\bar{u}_{0}(z)\vec{e}_{x}=\textit{Re}\,\bar{u}(z)\vec{e}_{x}\label{eq:v0}
\end{equation}
 for which (\ref{eq:NS}) reduces to 
\begin{equation}
\bar{u}''-\textit{Ha}^{2}\bar{u}=\bar{P},\label{eq:bflw}
\end{equation}
where $\textit{Re}=Uh/\nu$ is the Reynolds number based on the centreline
velocity $U$ of unperturbed flow, $\textit{Ha}=Bh\sqrt{\sigma/\rho\nu}$
is the Hartmann number, and $\bar{P}$ is a dimensionless coefficient
which relates $U$ with the applied pressure gradient $P=\bar{P}U\nu\rho/h^{2}$
by satisfying the normalization condition $\bar{u}(0)=1.$ This Reynolds
number is convenient for characterizing a flow driven by a fixed pressure
gradient. Besides this so-called pressure Reynolds number, one can
also use a Reynolds number based on the flow rate, $\textit{Re}_{q},$
which is more suitable for the case of fixed flow rate and will be
introduced in $\S$\ref{sec:Results}. Note that the use of either
of these two Reynolds numbers is a matter of convenience as long as
one remains a time-independent function of the other. Equation (\ref{eq:bflw})
defines the well-known Hartmann flow profile

\begin{equation}
\bar{u}(z)=\frac{\cosh(\textit{Ha})-\cosh(z\textit{Ha})}{\cosh(\textit{Ha})-1}\label{eq:Haflw}
\end{equation}
with $\bar{P}=-\frac{\textit{Ha}^{2}\cosh(\textit{Ha})}{\cosh(\textit{Ha})-1},$
which corresponds to a channel with perfectly conducting walls. In
a weak magnetic field $(\textit{Ha}\ll1),$ the Hartmann flow reduces
to the classic plane Poiseuille flow $\bar{u}(z)=1-z^{2}.$

Note that electrical conductivity of the walls affects only the relationship
between the applied pressure gradient and the centreline velocity
but not the profile of the Hartmann flow. Thus, the stability of Hartmann
flow with respect to transverse 2D disturbances, which are not affected
by the conductivity of the walls, is determined entirely by the centreline
velocity. This is the case considered in the present study.

\section{\label{sec:theory}Theoretical background}

\subsection{Linear stability of the base flow}

The two-dimensional travelling waves considered in this study are
expected to emerge as the result of linear instability of the Hartmann
flow (\ref{eq:v0}) with respect to infinitesimal perturbations $\vec{v}_{1}(\vec{x},t).$
Owing to the invariance of the base flow in both $t$ and $\vec{x}=(x,y),$
perturbations are sought as Fourier modes 
\begin{equation}
\vec{v}_{1}(\vec{r},t)=\vec{\hat{v}}(z)\e^{\lambda t+\mathrm{i}\vec{k}\cdot\vec{x}}+\mbox{c.c.}\label{eq:v1}
\end{equation}
defined by the complex amplitude distribution $\vec{\hat{v}}(z)$,
temporal growth rate $\lambda$ and the wave vector $\vec{k}=(\alpha,\beta).$
The incompressibility constraint, which takes the form $\vec{D}_{k}\cdot\vec{\hat{v}}=0,$
where $\vec{D}_{k}\equiv\vec{e}_{z}\frac{\mathrm{d}\,}{\mathrm{d}z}+\mathrm{i}\vec{k}$
is a spectral counterpart of the nabla operator, is satisfied by expressing
the component of the velocity perturbation in the direction of the
wave vector as $\hat{u}_{\shortparallel}=\vec{e}_{\shortparallel}\cdot\vec{\hat{v}}=\i k^{-1}\hat{w}',$
where $\vec{e}_{\shortparallel}=\vec{k}/k$ and $k=|\vec{k}|.$ Taking
the \emph{curl} of the linearized counterpart of (\ref{eq:NS}) to
eliminate the pressure gradient and then projecting it onto $\vec{e}_{z}\times\vec{e}_{\shortparallel},$
after some transformations we obtain a modified Orr-Sommerfeld-type
equation which includes a magnetic term 
\begin{equation}
\lambda\vec{D}_{k}^{2}\hat{w}=\left[\vec{D}_{k}^{4}-\textit{Ha}^{2}(\vec{e}_{z}\cdot\vec{D}_{k})^{2}+\mathrm{i}k\textit{Re}(\bar{u}''-\bar{u}\vec{D}_{k}^{2})\right]\hat{w}.\label{eq:OSm}
\end{equation}
The no-slip and impermeability boundary conditions require 
\begin{equation}
\hat{w}=\hat{w}'=0\quad\mbox{at}\quad z=\pm1.\label{eq:bc}
\end{equation}
The equation above is written in a non-standard form corresponding
to our choice of the characteristic velocity. Note that the Reynolds
number appears in this equation as a factor in the convective term
rather than its reciprocal in the viscous term as in the standard
form. As a result, the growth rate $\lambda$ differs by a factor
$\textit{Re}$ from its standard definition. 

Since equation (\ref{eq:OSm}) like its non-magnetic counterpart admits
Squire's transformation, in the following we consider only two-dimensional
perturbations $(k=\alpha),$ which are the most unstable \citep{Lock55}.
The problem is solved numerically using the Chebyshev collocation
method which is described in detail by \citet{Hagan-Priede-13a}.

\subsection{\label{sub:2D-states}2D nonlinear travelling waves}

Two-dimensional travelling waves emerge as follows. First, the neutrally
stable mode (\ref{eq:v1}) with a purely real frequency $\omega=-\mathrm{i}\lambda$
interacting with itself through the quadratically nonlinear term in
(\ref{eq:NS}) produces a steady streamwise-invariant perturbation
of the mean flow as well as a second harmonic $\sim\e^{2\mathrm{i}(\omega t+\alpha x)}.$
Further nonlinear interactions produce higher harmonics, which similarly
to the fundamental and second harmonic travel with the same phase
speed $c=-\omega/\alpha.$ Thus, the solution can be sought in the
form
\begin{equation}
\vec{v}(\vec{r},t)=\sum_{n=-\infty}^{\infty}E^{n}\vec{\hat{v}}_{n}(z),\label{eq:v2d}
\end{equation}
where $E=\e^{\mathrm{i}(\omega t+\alpha x)}$ contains $\omega,$
which needs to determined together $\vec{\hat{v}}_{n}$ by solving
a nonlinear eigenvalue problem \citep{Huerre-Rossi-98}. The reality
of solution requires $\vec{\hat{v}}_{-n}=\vec{\hat{v}}_{n}^{*},$
where the asterisk stands for the complex conjugate. The incompressibility
constraint applied to the $n$th velocity harmonic results in $\vec{D}_{\alpha_{n}}\cdot\vec{\hat{v}}_{n}=0,$
where $\vec{D}_{\alpha_{n}}\equiv\vec{e}_{z}\frac{\mathrm{d}\,}{\mathrm{d}z}+\mathrm{i}\vec{e}_{x}\alpha_{n}$
with $\alpha_{n}=\alpha n$ stands for the spectral counterpart of
the nabla operator. This constraint can be satisfied by expressing
the streamwise velocity component 
\begin{equation}
\hat{u}_{n}=\vec{e}_{x}\cdot\vec{\hat{v}}_{n}=\mathrm{i}\alpha_{n}^{-1}\hat{w}_{n}'\label{eq:un}
\end{equation}
in terms of the transverse component $\hat{w}_{n}=\vec{e}_{z}\cdot\vec{\hat{v}}_{n},$
which we employ instead of the commonly used stream function. Henceforth,
the prime is used as a shorthand for $d/dz.$ Note that (\ref{eq:un})
is not applicable to the zeroth harmonic, for which it yields $\hat{w}_{0}\equiv0.$
Thus, $\hat{u}_{0}$ needs to be considered separately in this velocity-based
formulation. 

Taking the \emph{curl} of (\ref{eq:NS}) to eliminate the pressure
gradient and then projecting it onto $\vec{e}_{y},$ we obtain 
\begin{equation}
[\vec{D}_{\alpha_{n}}^{2}-\mathrm{i}\omega n]\hat{\zeta}_{n}-\textit{Ha}^{2}\hat{u}_{n}'=\hat{h}_{n},\label{eq:vrt}
\end{equation}
where 
\begin{equation}
\hat{\zeta}_{n}=\vec{e}_{y}\cdot\vec{D}_{\alpha_{n}}\times\vec{\hat{v}}_{n}=\begin{cases}
\mathrm{i}\alpha_{n}^{-1}\vec{D}_{\alpha_{n}}^{2}\hat{w}_{n}, & n\not=0;\\
\hat{u}_{0}', & n=0.
\end{cases}\label{eq:zeta}
\end{equation}
and 
\begin{equation}
\hat{h}_{n}={\displaystyle \sum_{m}}\vec{\hat{v}}_{n-m}\cdot\vec{D}_{\alpha_{m}}\hat{\zeta}_{m}\label{eq:h}
\end{equation}
are the $y$-components of the $n$th harmonic of the vorticity $\vec{\zeta}=\vec{\nabla}\times\vec{v}$
and that of the \emph{curl} of the nonlinear term $\vec{h}=\vec{\nabla}\times(\vec{v}\cdot\vec{\nabla})\vec{v}.$
Henceforth, the omitted summation limits are assumed to be infinite.
Separating the terms involving $\hat{u}_{0}$ in (\ref{eq:h}), it
can be rewritten as $\hat{h}_{n}=\mathrm{i}\alpha_{n}^{-1}(\hat{h}_{n}^{w}+\hat{h}_{n}^{u}),$
where 
\begin{eqnarray}
\hat{h}_{n}^{w} & = & n{\displaystyle \sum_{m\not=0}m^{-1}(}\hat{w}_{n-m}\vec{D}_{\alpha_{m}}^{2}\hat{w}_{m}'-\hat{w}_{m}'\vec{D}_{\alpha_{n-m}}^{2}\hat{w}_{n-m}),\label{eq:hw}\\
\hat{h}_{n}^{u} & = & \mathrm{i}\alpha_{n}[\hat{u}_{0}-\hat{u}_{0}''\vec{D}_{\alpha_{n}}^{2}]\hat{w}_{n}\equiv\mathcal{N}_{n}(\hat{u}_{0})\hat{w}_{n}.\label{eq:hu}
\end{eqnarray}
 Eventually, using the expressions above, (\ref{eq:vrt}) can be written
as 
\begin{equation}
\mathcal{L}_{n}(\mathrm{i}\omega,\hat{u}_{0})\hat{w}_{n}=\hat{h}_{n}^{w},\label{eq:wn}
\end{equation}
with the operator 
\begin{equation}
\mathcal{L}_{n}(\mathrm{i}\omega,\hat{u}_{0})=[\vec{D}_{\alpha_{n}}^{2}-\mathrm{i}\omega n]\vec{D}_{\alpha_{n}}^{2}-\textit{Ha}^{2}(\vec{e}_{z}\cdot\vec{D}_{\alpha_{n}})^{2}-\mathcal{N}_{n}(\hat{u}_{0}).\label{eq:Ln}
\end{equation}
This equation governs all harmonics except the zeroth one, for which,
in accordance with the incompressibility constraint (\ref{eq:un}),
it implies $\hat{w}_{0}\equiv0$. The zeroth velocity harmonic, which
has only the streamwise component $\hat{u}_{0},$ is governed directly
by the $x$-component of the Navier--Stokes equation (\ref{eq:NS}):
\begin{equation}
\hat{u}_{0}''-\textit{Ha}^{2}\hat{u}_{0}=\hat{P}_{0}+\hat{g}_{0},\label{eq:u0}
\end{equation}
where $\hat{P}_{0}=\bar{P}\textit{Re}$ is a dimensionless mean pressure
gradient and 
\begin{equation}
\hat{g}_{0}=\mathrm{i}\sum_{m\not=0}\alpha_{m}^{-1}(\hat{w}_{m}^{*}\hat{w}_{m}')'\label{eq:g0}
\end{equation}
is the $x$-component of the zeroth harmonic of the nonlinear term
$\vec{g}=(\vec{v}\cdot\vec{\nabla})\vec{v}.$ Velocity harmonics are
subject to the usual no-slip and impermeability boundary conditions

\begin{equation}
\hat{w}_{n}=\hat{w}_{n}'=\hat{u}_{0}=0\textrm{ at }z=\pm1.\label{eq:bc-1}
\end{equation}

\subsection{\label{sub:2Dlstab}Linear stability of 2D travelling waves}

Weakly subcritical equilibrium states, which exist in this case \citep{Hagan-Priede-13b},
are unconditionally unstable \citep{Schmid-Henningson-01}. This is
because the growth rate of subcritical disturbances increases with
their amplitude \citep{Hagan-Priede-13b}. Thus, a disturbance with
an amplitude slightly lower or higher than the equilibrium one will
respectively decay or grow so diverging from the equilibrium state.
The stability of strongly subcritical equilibrium states is not obvious.
\citet{Orszag-Patera-83} originally suggested that the subcritical
equilibrium state appearing at the linear stability threshold remains
linearly unstable down the lowest possible Reynolds number admitting
such states. At this limiting Reynolds number, which is the main concern
of the present study, the unstable subcritical state disappears by
merging with another equilibrium state of a higher amplitude. The
latter was thought by \citet{Orszag-Patera-83} to be linearly stable
as in the saddle--node bifurcation. This simple picture was amended
by \citet{Pugh-Saffman88} who showed that this is the case when the
flow is driven by a fixed flow rate but not by a fixed pressure gradient.
Although travelling-wave states at fixed flow rate are physically
equivalent to those at fixed pressure gradient \citep{Soibelman-Meiron-91},
it is not the case in general when the flow rate and the mean pressure
gradient depend not only on each other but also on time. First of
all, the distinction between flows driven by fixed pressure gradient
and fixed flow rate becomes important when the stability of travelling-wave
states is considered.

Linear stability of the travelling-wave states, which in contrast
to the rectilinear base state are periodic rather than invariant in
both the time and the streamwise direction, is described by Floquet
theory \citep{Bender-Orsag78,Herb88} according to which a small-amplitude
velocity disturbance can be sought similarly to (\ref{eq:v2d}) as
\begin{equation}
\vec{v}_{1}(\vec{r},t)=e^{\tilde{\lambda}t}\sum_{n=-\infty}^{\infty}E^{n+\epsilon}\vec{\tilde{v}}_{n}(z)+\mbox{c.c.},\label{eq:v1-flq}
\end{equation}
where $\tilde{\lambda}$ is generally a complex growth rate and $\epsilon$
is a real detuning parameter which defines the sideband wavenumber
$\tilde{\alpha}=\epsilon\alpha$ \citep{Huerre-Rossi-98}, also called
the subharmonic wavenumber by \citet{Soibelman-Meiron-91}. The case
$\epsilon=0$ corresponds to the fundamental mode, which is also called
superharmonic \citep{Pugh-Saffman88,Soibelman-Meiron-91}, whereas
$\epsilon=\pm\frac{1}{2}$ correspond to the so-called subharmonic
mode. The disturbances with other values of $\epsilon$ are referred
to as combination or detuned modes. Note that (\ref{eq:v1-flq}) with
$\epsilon\pm1$ is equivalent to the mode with $\epsilon$ and the
index $n$ shifted by one. In addition, as seen from (\ref{eq:v1-flq})
the modes with $\pm\epsilon$ are complex conjugate. Thus, it suffices
to consider $0\le\epsilon<1$ or any other interval of $\epsilon$
of length one. Separating the two modes with opposite $z$-symmetries,
as discussed in the next section, this interval can be reduced by
half.

Adding disturbance (\ref{eq:v1-flq}) to the travelling-wave base
state (\ref{eq:v2d}), we obtain a linearized counterpart of (\ref{eq:vrt})
for the transverse velocity harmonic $\tilde{w}_{n}:$

\begin{equation}
\left[\vec{D}_{\tilde{\alpha}_{n}}^{2}-\tilde{\lambda}-\mathrm{i}\omega n\right]\vec{D}_{\tilde{\alpha}_{n}}^{2}\tilde{w}_{n}-\textit{Ha}^{2}\tilde{w}_{n}''=-\mathrm{i}\tilde{\alpha}_{n}\sum_{m}\vec{D}_{\tilde{\alpha}_{n}}\cdot(\vec{\hat{v}}_{n-m}\tilde{\zeta}_{m}+\vec{\tilde{v}}_{m}\hat{\zeta}{}_{n-m}),\label{eq:flq}
\end{equation}
where $\tilde{\alpha}_{n}=\alpha_{n}+\tilde{\alpha}=(n+\epsilon)\alpha$
is a detuned wavenumber for the $n$th harmonic, $\tilde{u}_{n}=\vec{e}_{x}\cdot\vec{\tilde{v}}_{n}=\mathrm{i}\tilde{\alpha}_{n}^{-1}\tilde{w}_{n}'$
is the streamwise velocity and $\tilde{\zeta}_{n}=\mathrm{i}\tilde{\alpha}_{n}^{-1}\vec{D}_{\tilde{\alpha}_{n}}^{2}\tilde{w}_{n}$
is the spanwise vorticity of the disturbance. The boundary conditions,
as usual, are $\tilde{w}_{n}(1)=\tilde{w}_{n}'(1)=0.$ 

Cutting the series (\ref{eq:v1-flq}) off at $n=\pm N,$ we obtain
a linear eigenvalue problem represented by $2N+1$ complex equations
(\ref{eq:flq}) for the eigenvector consisting of the same number
of harmonics $\tilde{w}_{n}$ with the eigenvalue $\tilde{\lambda},$
which depends on the subharmonic wavenumber $\tilde{\alpha}.$ This
eigenvalue problem is solved in the same way as that for the linear
stability of the rectilinear base flow. To avoid the division by zero
in the expressions for $\tilde{u}_{0}$ and $\tilde{\zeta}_{0}$ above,
which occurs for $\tilde{\alpha}=0,$ we use the substitution $\tilde{w}_{n}=-\mathrm{i}\tilde{\alpha}_{n}\tilde{\psi}_{n},$
where $\tilde{\psi}_{n}$ is the stream function. This makes superharmonic
disturbances treatable in the same way as detuned ones. For the superharmonic
disturbances corresponding to $\tilde{\alpha}\rightarrow0,$ the eigenvalue
problem becomes self-adjoint with $\tilde{w}_{-n}=\tilde{w}_{n}^{*},$
which means that eigenvalues are either real or complex conjugate. 

Note that the detuned disturbances with $\tilde{\alpha}\not=0$ affect
neither the mean pressure gradient nor the flow rate. Both of these
quantities are associated with a zero-wavenumber mode, which occurs
only for the fundamental disturbances $\epsilon=0$ when the zeroth
harmonic of the streamwise velocity perturbation is an odd function
of $z,$ i.e., $\tilde{u}_{0}(-z)=-\tilde{u}_{0}(z).$ This is the
case for the fundamental modes with $z$-parities opposite to those
of the travelling-wave state which are discussed in the next section.

The problem posed by (\ref{eq:flq}) corresponds to the case of fixed
flow rate. For $\tilde{\alpha}_{0}=0$ we have $\vec{D}_{\tilde{\alpha}_{0}}\equiv\vec{e}_{z}\frac{d\,}{dz}$
and thus for $n=0$ equation (\ref{eq:flq}) can be integrated once.
This results in a linearized counterpart of (\ref{eq:u0}) for $\tilde{\psi}'_{0}=\tilde{u}_{0}$
containing a constant of integration $\tilde{P}_{0}$ which represents
perturbation of the mean pressure gradient:
\begin{equation}
\tilde{u}_{0}''-(\tilde{\lambda}+\textit{Ha}^{2})\tilde{u}_{0}=\sum_{m}(\hat{w}_{m}^{*}\tilde{u}_{m}+\hat{u}{}_{m}^{*}\tilde{w}_{m})'+\tilde{P}_{0}.\label{eq:ut0}
\end{equation}
Using this equation with $\tilde{P}_{0}=0$ instead of the original
one we obtain a linear stability problem for the travelling waves
driven by a fixed mean pressure gradient. Equation (\ref{eq:ut0})
can integrated once more leading to an equivalent equation in terms
of $\tilde{\psi}{}_{0}$ which can be used instead (\ref{eq:ut0}).
Alternatively, the integrated equation can be used as an effective
but rather complicated boundary condition replacing the fixed flow
rate condition $\tilde{\psi}_{0}(1)=0$ in the original equation (\ref{eq:flq})
for $\tilde{\psi}_{0}$ \citep{Soibelman-Meiron-91}.

\section{\label{sec:Num}Numerical method}

The problem is solved numerically using a Chebyshev collocation method
with the Chebyshev-Lobatto nodes 
\begin{equation}
z_{i}=\cos\left(i\pi/2M\right),\quad i=0,\cdots,M,\label{zcol}
\end{equation}
at which the discretized solution $(\hat{w}_{n},\hat{u}_{0})(z_{i})=(\tilde{w}_{n},\tilde{u}_{0})_{i}$
is sought in the upper half of the channel. The reduction to half
the channel is due to the following symmetries implied by (\ref{eq:hw}),
which is the nonlinear term of the vorticity equation (\ref{eq:vrt}).
The $z$-symmetry of this term is the same of as that of $\hat{w}_{n}$
governed by (\ref{eq:vrt}), and it is determined by the first derivative
of the terms quadratic in $\hat{w}_{n}.$ Note that the first derivative
inverts $z$-symmetry and the second derivative  conserves it. Since
the product of two harmonics with indices $m$ and $n$ produces two
harmonics with the same $z$-symmetry and indices $n\pm m,$ which
are separated by an even number $2m$, all harmonics with the same
parity of indices have the same $z$-symmetry. Harmonics with even
indices are produced by the products of harmonics with  indices of
the same parity, which have the same $z$-symmetry. Thus, these products
are always even functions of $z.$ The inversion of the $z$-symmetry
by the first derivative in the nonlinear term results in $\hat{w}_{n}$
with even $n$ being an odd function of $z.$ All harmonics with odd
indices can have either odd or even $z$-symmetry depending on the
symmetry of the fundamental mode $(n=1),$ which is usually determined
by the linear stability analysis. This gives rise to two types of
possible 2D solutions satisfying $\hat{w}_{n}(-z)=(-1)^{n+1}\hat{w}_{n}(z)$
and $\hat{w}_{n}(-z)=-\hat{w}_{n}(z)$, which are referred to as even
or odd depending on the $z$-symmetry of the transverse velocity harmonics
with odd indices. In secondary stability analysis, small-amplitude
perturbations of each of these two solution branches splits into two
further symmetry types. The first type of possible perturbation consists
of the harmonics with the same $z$-parities as the travelling-wave
solution itself, i.e. $\tilde{w}_{n}(-z)=(-1)^{n+1}\tilde{w}_{n}(z)$
and $\tilde{w}_{n}(-z)=-\tilde{w}_{n}(z)$ for even and odd branches,
respectively. The second type of perturbations admitted by (\ref{eq:flq})
consists of harmonics with $z$-symmetries opposite to those of the
travelling-wave solution, i.e. $\tilde{w}_{n}(-z)=(-1)^{n}\tilde{w}_{n}(z)$
and $\tilde{w}_{n}(-z)=\tilde{w}_{n}(z)$ for even and odd branches,
respectively \citep{Pugh-Saffman88}. Note that for the even 2D solution
branch, which consists of harmonics with alternating $z$-parities,
one type of perturbation is changed into the other by the transformation
$\epsilon\rightarrow\epsilon\pm1$ which effectively shifts the index
of harmonics in (\ref{eq:v1-flq}) by one.

Equations are approximated at the internal collocation points $1\leq i\leq M$
by using differentiation matrices which express the derivatives in
terms of the collocation variables $(\underline{w}_{n},\underline{u}_{0})_{i}.$
Odd or even $z$-symmetry of each particular harmonic is taken into
account by the reduced-size differentiation matrices based on the
collocation points only in one half of the channel. The boundary conditions
(\ref{eq:bc}) are imposed at the boundary point $i=0$ \citep{Peyret-02}.
When series (\ref{eq:v2d}) are truncated at $n=\pm N,$ (\ref{eq:wn})
reduce to $N\times M$ complex algebraic equations with respect to
the same number of complex unknowns $\underline{w}_{n}$ for $n=1,2,\ldots,N.$
Note that $\underline{w}_{0}\equiv0$ and $\underline{w}_{-n}=\underline{w}_{n}^{*}.$
The equations also contain $M$ real unknowns $\tilde{u}_{0},$ which
are governed by the same number of real equations resulting from the
collocation approximation of (\ref{eq:u0}). The equations are linear
and, thus, directly solvable for $\tilde{u}_{0}$ in terms of $\underline{w}_{n}$
by inverting the respective matrices. This leads to a system of $N\times M$
nonlinear complex equations for the same number of complex unknowns
$\underline{w}_{n}.$ Since the equations also contain $\underline{w}_{n}^{*},$
the actual unknowns are the real and imaginary parts of $\underline{w}_{n},$
which need to be determined by solving the same number, i.e. $2N\times M,$
of real equations represented by the real and imaginary parts of the
original complex equations. 

However, there is one more unknown: the frequency $\omega,$ which
is the eigenvalue of this nonlinear problem, and needs to be determined
along with $\underline{w}_{n}.$ There are two ways to balance the
number of unknowns and equations. First, as the problem is homogeneous,
non-trivial solution requires a solvability condition to be satisfied,
which provides another equation analogous to the characteristic determinant
in the case of a linear eigenvalue problem. The second possibility,
which is used here, follows from the fact that owing to the translational
invariance of the problem $\underline{w}_{n}$ is defined up an arbitrary
phase. The phase, which determines the $x$-offset of the wave, can
be fixed by imposing the condition $\Im[\underline{w}_{1,i}]=0$ at
some collocation point $i$ where the solution is not already fixed
by boundary or symmetry conditions. This is equivalent to setting
$\underline{w}_{1,i}=A,$ where $A$ is a real parameter defining
the amplitude of the transverse velocity. Thus, the number of unknowns
is reduced by one and the system of $2N\times M$ nonlinear algebraic
equations can be written in the general form 
\begin{equation}
\underline{F}(\underline{w},A,\omega;\alpha,\textit{Re})\underline{w}=\underline{0},\label{eq:nlin}
\end{equation}
 where $\underline{w}$ are the real and imaginary parts of $\underline{w}_{n}$
normalized with the real amplitude $A$ and $\underline{F}$ is a
real square matrix of size $2N\times M$ depending on the listed parameters.
For given $\alpha$ and $\textit{Re},$ this problem can be solved
by the Newton-Raphson method with respect to $A,$ $\omega$ and $2N\times M-2$
unknown $\underline{w}.$ In some cases, instead of $\textit{Re},$
it is more convenient to fix $A$ and then to solve for $\textit{Re}$
depending on $A$ and $\alpha.$ 

The solution is traced using a quadratic extrapolation along the arclength
in logarithmic coordinates. For a general function $f(p)$ of an argument
$p$ the scale-independent arclength element is defined as $\delta s^{2}=\ln^{2}(1+\frac{\delta f}{f})+\ln^{2}(1+\frac{\delta p}{p}).$
Starting with a reference arclength based on the solution at the first
three chosen parameter values, a subsequent parameter value and an
initial guess for the Newton-Raphson method are extrapolated from
the previous three values. When Newton-Raphson iterations fail, the
step size along the arclength is reduced until a solution is recovered,
and then gradually increased to its original value as the solution
is successfully traced.

\section{\label{sec:Results}Results}

\subsection{Nonlinear 2D travelling waves}

\begin{table}
\begin{centering}
\begin{tabular}{cccccc}
$M\times N$ & $\textit{Re}_{n}$ & $\alpha_{n}$ & $c_{n}=-\omega/\textit{Re}_{n}\alpha$ & $A_{E}^{2}(\times10^{3})$ & $A=\hat{w}_{1}(0)$\tabularnewline
$32\times1$ & $2825.56$ & $1.22223$ & $0.345828$ & $6.14777$ & $131.655$\tabularnewline
$32\times2$ & $2701.72$ & $1.31294$ & $0.366290$ & $4.92982$ & $118.206$\tabularnewline
$32\times3$ & $2911.36$ & $1.31824$ & $0.364025$ & $4.33693$ & $119.120$\tabularnewline
$32\times4$ & $2933.53$ & $1.32425$ & $0.364470$ & $4.50019$ & $121.571$\tabularnewline
$32\times6$ & $2940.08$ & $1.31701$ & $0.363147$ & $4.26584$ & $119.171$\tabularnewline
$32\times8$ & $2939.05$ & $1.31752$ & $0.363251$ & $4.28277$ & $119.330$\tabularnewline
$32\times10$ & $2939.04$ & $1.31751$ & $0.363250$ & $4.28224$ & $119.324$\tabularnewline
$40\times10$ & $2939.04$ & $1.31750$ & $0.363249$ & $4.28224$ & $119.320$\tabularnewline
\end{tabular}
\par\end{centering}

\protect\caption{\label{tab:PPF}Critical parameters for the appearance of 2D travelling
waves in plane Poiseuille flow computed with various number of collocation
points $M$ and harmonics $N.$}
\end{table}

Weakly nonlinear analysis shows that the instability of the Hartmann
flow is invariably subcritical regardless of the magnetic field strength
\citep{Hagan-Priede-13b}. In the present study, we determine how
far the subcritical equilibrium states, which bifurcate from the Hartmann
flow, extend below the linear stability threshold. Let us first validate
our method described in Sec. \ref{sub:2D-states} by computing the
critical Reynolds number for 2D travelling waves in plane Poiseuille
flow, which corresponds to $\textit{Ha}=0.$ By solving (\ref{eq:nlin})
for $\textit{Re}$ as a function of $A$ and $\alpha$ and then minimizing
the solution over both variables we obtain the critical values which
are shown in table \ref{tab:PPF} for various numbers of collocation
points $M$ and harmonics $N.$ The critical parameters for the first
three numerical resolutions perfectly agree with those found by \citet{Soibelman-Meiron-91},
whereas for the last three resolutions both $\textit{Re}$ and $\alpha$
agree up to 5 decimal points with the accurate results obtained by
\citet{Casas-Jorba-12} using $2M=70$ Chebyshev polynomials and $N=22$
Fourier modes. To characterize the deviation of the velocity distribution
(\ref{eq:v2d}) from the base state (\ref{eq:v0}), besides the transverse
velocity amplitude $A$ introduced in (\ref{eq:nlin}), we also use
the amplitude associated with the energy of perturbation scaled by
the energy of the basic flow: 
\begin{equation}
A_{E}^{2}=\int_{0}^{1}\left\langle |\vec{v}(x,z)-\vec{v}_{0}(z)|^{2}\right\rangle dz/\int_{0}^{1}|\vec{v}_{0}(z)|^{2}dz,\label{eq:nrg}
\end{equation}
where the angle brackets stand for the streamwise average. This quantity
slightly differs from that used by \citet{Soibelman-Meiron-91} who
neglect the contribution of the mean flow perturbation. 

\begin{figure}
\begin{centering}
\includegraphics[width=0.5\textwidth]{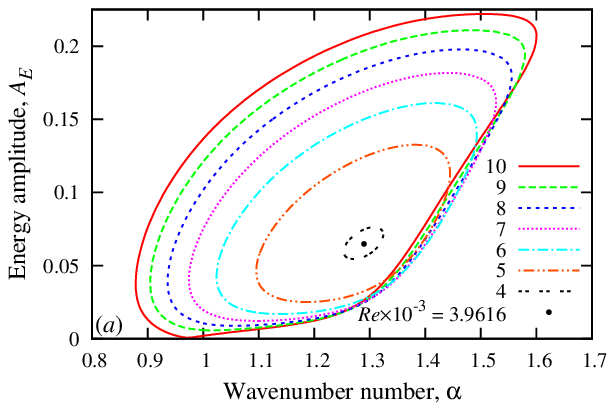}\includegraphics[width=0.5\textwidth]{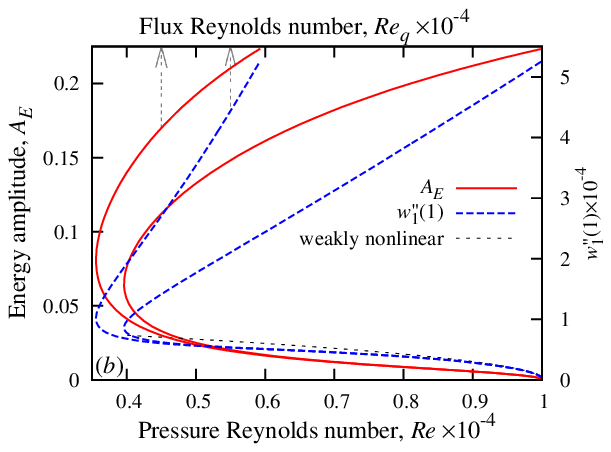}
\par\end{centering}

\protect\caption{\label{fig:erg-Ha1}The energy amplitude of equilibrium states of
the even mode versus the wavenumber $\alpha$ at various $\textit{Re}$
(a) and the extrema of the energy amplitude versus pressure Reynolds
number $\textit{Re}$ (bottom axis) and versus the flux Reynolds number
(top axis) (b) for $\textit{Ha}=1$ computed with the resolution $M\times N=32\times8.$ }
\end{figure}

\begin{figure}
\begin{centering}
\includegraphics[width=0.5\textwidth]{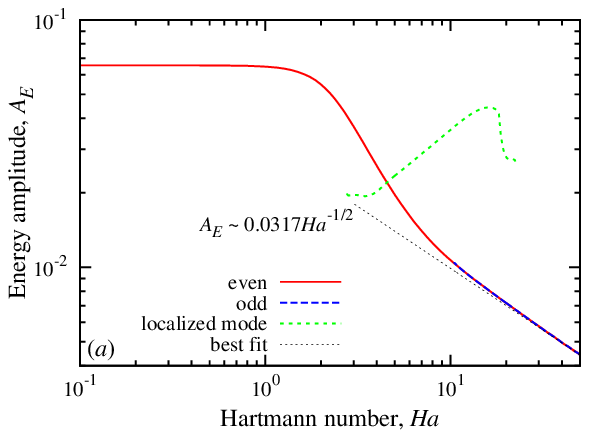}\includegraphics[width=0.5\textwidth]{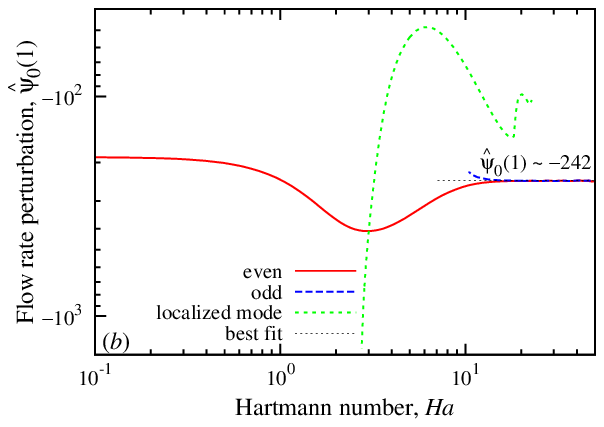}
\par\end{centering}

\protect\caption{\label{fig:nln2}Energy amplitude (a) and the flow rate perturbation
(b) at the 2D nonlinear instability threshold for even and odd modes
versus the Hartmann number. }
\end{figure}

We start with a relatively low Hartmann number $\textit{Ha}=1$ for
which the flow becomes linearly unstable at $\textit{Re}_{l}=10016.3$
with respect to a mode with $\alpha_{l}=0.971827$ and even $z$-symmetry
as in the non-magnetic case. The energy amplitude of equilibrium states
versus the wavenumber is plotted in figure \ref{fig:erg-Ha1}(a) for
various subcritical values of $\textit{Re}.$ As for the non-magnetic
plane Poiseuille flow, equilibrium states form closed contours, which
shrink as $\textit{Re}$ is reduced, and collapse to a point at the
critical $\textit{Re}_{n}=3961.36$ below which 2D travelling waves
vanish. It means that subcritical perturbations have both a lower
and an upper equilibrium amplitude. Both these amplitudes are plotted
in figure \ref{fig:erg-Ha1}(b) together with the respective value
of $\hat{w}_{1}''(1),$ which is the quantity predicted by the weakly
nonlinear analysis \citep{Hagan-Priede-13b}, versus the usual pressure
Reynolds number on the bottom axis and versus the flux Reynolds number
$\textit{Re}_{q}$ on the top axis. The latter is related to the original
Reynolds number $\textit{Re}$ based on the mean pressure gradient:
\begin{equation}
\textit{Re}_{q}=\textit{Re}+\hat{\psi}_{0}(1)/\bar{\psi}(1),\label{eq:Req}
\end{equation}
where $\bar{\psi}(1)=\int_{0}^{1}\bar{u}(z)\, dz=(\cosh(\textit{Ha})-\textit{Ha}^{-1}\sinh(\textit{Ha}))/(\cosh(\textit{Ha})-1)$
is half of the flux carried by unperturbed Hartmann flow (\ref{eq:Haflw})
and $\hat{\psi}_{0}(1)$ is the flow rate perturbation defined by
(\ref{eq:psi0}) and plotted in figure \ref{fig:nln2}(b) below. As
seen, the lower branch of $\hat{w}_{1}''(1)$ is predicted well by
the weakly nonlinear solution for subcritical Reynolds numbers down
to $\textit{Re}\approx7000.$

\begin{figure}
\begin{centering}
\includegraphics[width=0.5\textwidth]{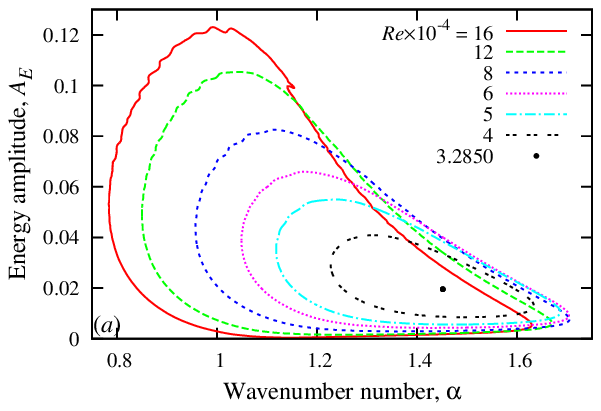}\includegraphics[width=0.5\textwidth]{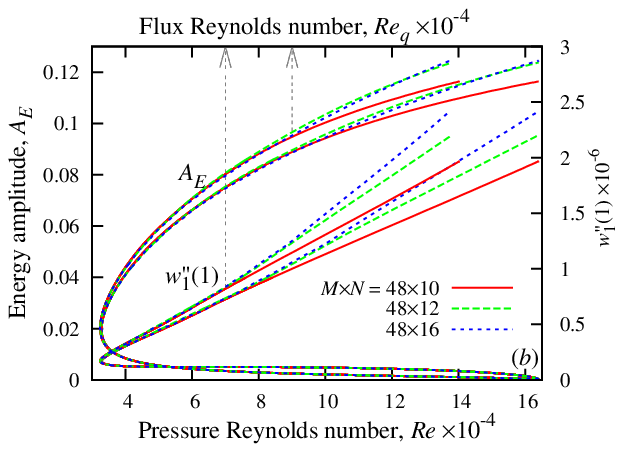}
\par\end{centering}

\protect\caption{\label{fig:erg-Ha5}The energy amplitude of equilibrium states of
the even mode versus the wavenumber $\alpha$ at various $\textit{Re}$
(a) and the extrema of the energy amplitude versus $\textit{Re}$
(b) for $\textit{Ha}=5$ computed with the resolutions $M\times N=48\times16\cdots32.$ }
\end{figure}

A similar structure of subcritical equilibrium states is also seen
in figure \ref{fig:erg-Ha5} for $\textit{Ha}=5$ when the flow becomes
linearly unstable at $\textit{Re}_{l}=164\thinspace154$ with respect
to a perturbation of even $z$-symmetry. At this large $\textit{Re}$
it becomes difficult to compute accurately the upper equilibrium states
which remain wiggly up to the numerical resolution of $48\times32.$
Strongly subcritical states, which in this case extend down to $\textit{Re}_{n}\approx32\thinspace860,$
can reliably be computed with a substantially lower resolution of
$48\times16.$ This structure of the equilibrium states of the even
mode is typical in the vicinity of the critical Reynolds number at
higher Hartmann numbers also. In the following, we focus on such strongly
subcritical Reynolds numbers at which 2D travelling waves emerge.
The respective Reynolds number defines the 2D nonlinear stability
threshold.

\begin{figure}
\begin{centering}
\includegraphics[width=0.5\textwidth]{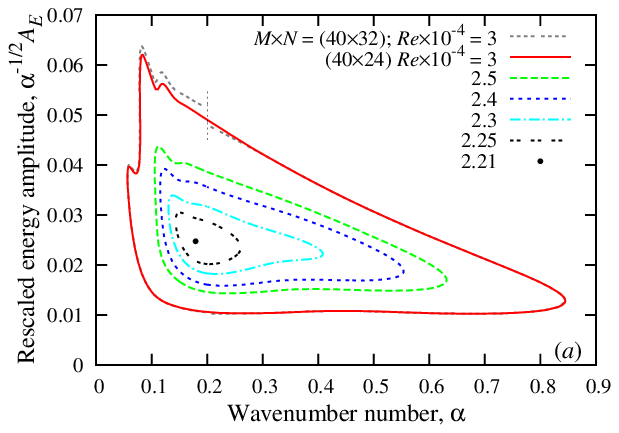}\includegraphics[width=0.5\textwidth]{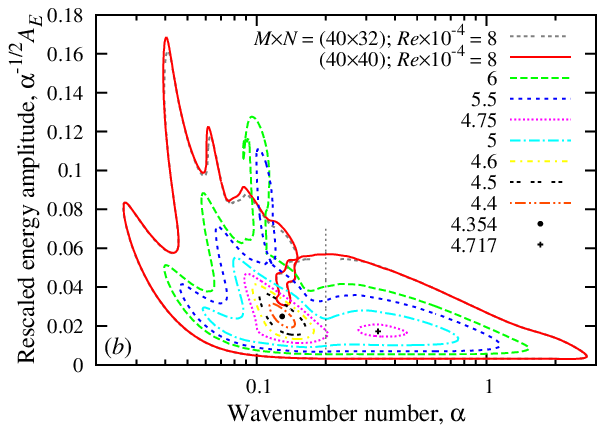}
\par\end{centering}

\protect\caption{\label{fig:erg-odd}The energy amplitude of equilibrium states of
the odd mode rescaled with $\alpha^{-1/2}$ versus the wavenumber
$\alpha$ for $\textit{Ha}=5$ (a), $\textit{Ha}=10$ (b) and various
$\textit{Re}$ close to the turning point. The wavenumbers $\alpha<0.2$
for $Re=3\times10^{4},$ $N=32$ (a) and $Re=8\times10^{4},$ $N=40$
(b) is rescaled, respectively with factors $32/24$ and $40/32.$}
\end{figure}

\begin{figure}
\begin{centering}
\includegraphics[bb=0bp 0bp 227bp 154bp,clip,width=0.5\textwidth]{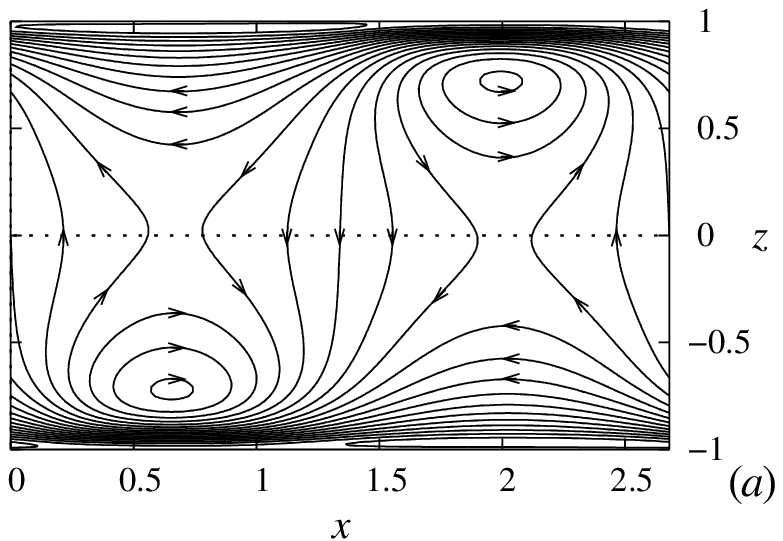}\includegraphics[bb=0bp 0bp 227bp 154bp,clip,width=0.5\textwidth]{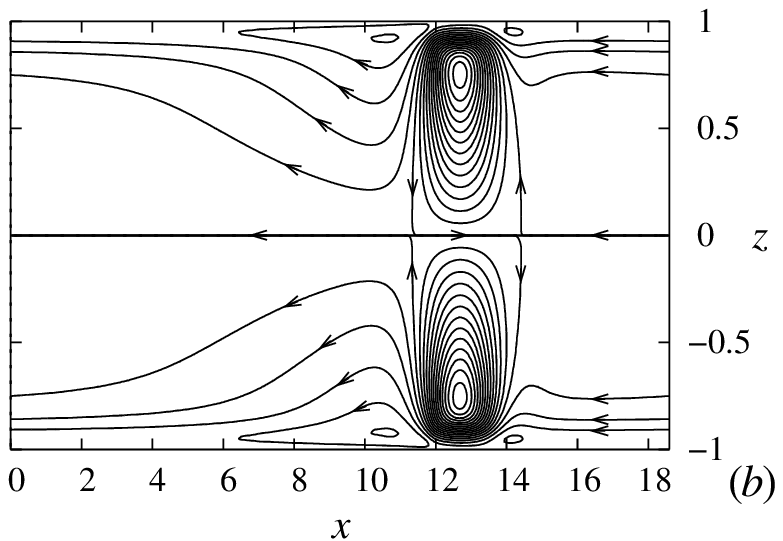}
\par\end{centering}

\begin{centering}
\includegraphics[bb=0bp 0bp 226bp 154bp,clip,width=0.5\textwidth]{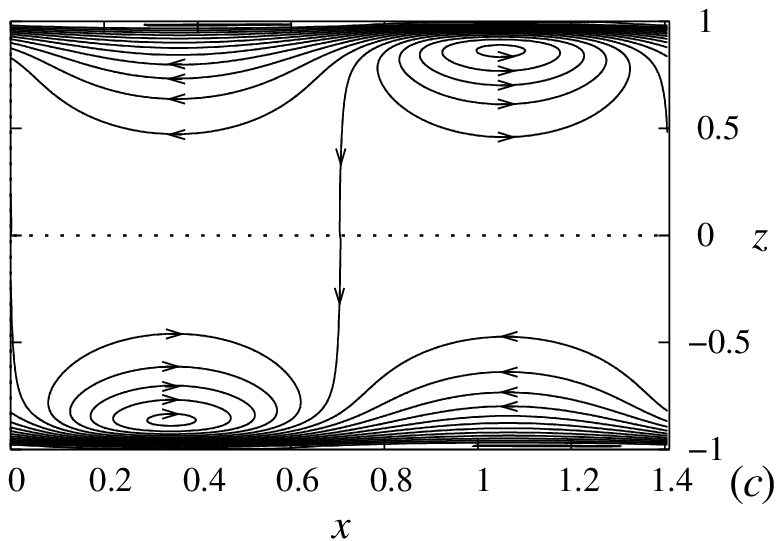}\includegraphics[bb=0bp 0bp 227bp 154bp,clip,width=0.5\textwidth]{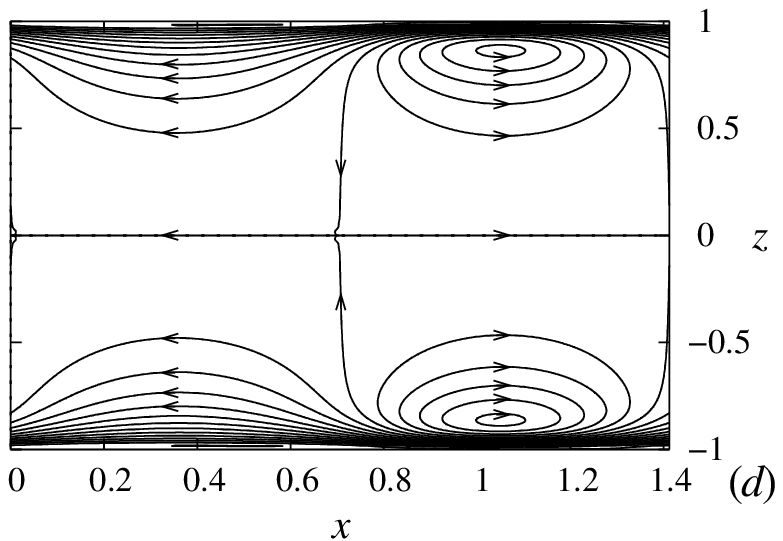}
\par\end{centering}

\protect\caption{\label{fig:str2}Streamlines of even (a, c) and odd (b, d) critical
finite-amplitude perturbations for $\textit{Ha}=10$ (a, b) and $\textit{Ha}=20$
(c, d). }
\end{figure}

In contrast to Poiseuille flow, which can be linearly unstable only
to perturbations of even $z$-symmetry, Hartmann flow can be linearly
unstable also to perturbations of odd $z$-symmetry when $Ha\gtrsim6.5$
\citep{Hagan-Priede-13b}. Subcritical equilibrium states of the odd
mode, whose energy amplitudes are shown in figure \ref{fig:erg-odd}
versus the wavenumber for $\textit{Ha}=5$ and $\textit{Ha}=10$ at
several Reynolds numbers in the vicinity of the critical point, have
a significantly different structure. Besides multiple minima and loopy
structures, which are seen to form at $\textit{Ha}=10$ in figure
\ref{fig:erg-odd}(b), the main difference from the even mode is the
extension of these states towards small wavenumbers. These apparently
long-wave states have several numerical peculiarities. First, when
the number of harmonics $N$ is increased, the wavelength of these
states increases whereas their amplitude decreases. At the same time,
the states with sufficiently short wavelength $(\alpha\gtrsim0.2)$
converge. Although long-wave states do not appear to converge, the
pattern of their equilibrium energy amplitude becomes self-similar
at sufficiently large $N.$ Both the wavenumber $\alpha$ and the
energy (\ref{eq:nrg}) for these states decrease inversely with $N.$
This is seen in figure \ref{fig:erg-odd}(a) where the rescaled energy
amplitude $\alpha^{-1/2}A_{E}$ computed for $Re=3\times10^{4}$ with
$N=32$ harmonics overlaps with the respective amplitude computed
with $N=24$ harmonics when the range of wavenumbers $\alpha<0.2$
is rescaled with a factor of $\frac{32}{24}$. Note that the same
results overlap without rescaling at larger $\alpha.$ Similar behaviour
can be seen also in figure \ref{fig:erg-odd}(b) where a much more
complicated equilibrium amplitude distribution for $Re=8\times10^{4}$
computed with $N=32$ closely overlaps with the respective distribution
computed with $N=40$ when $\alpha<0.2$ is rescaled with a factor
of $\frac{40}{32}.$ This rescaling has two implications. First, the
relevant parameter for these apparently long-wave equilibrium states
is not the wavenumber of the first harmonic $\alpha$ but that of
the last harmonic given by$\alpha N.$ Second, these states are characterized
by the integral perturbation energy over the wavelength $\propto\alpha^{-1}A_{E}^{2}$
rather by its streamwise average (\ref{eq:nrg}). These peculiar properties
of the long-wave equilibrium states are due their unusual spatial
structure which is revealed by the streamlines of the critical perturbation
shown for $Ha=10$ in figure \ref{fig:str2}(b) below. Namely, these
states turn out be localized rather than wave-like. Periodicity of
these states is enforced by the Fourier series representation (\ref{eq:v2d}).
The apparent period is determined by the fundamental wavenumber $\alpha$
which decreases inversely with the number of harmonics. At the same
time, the actual solution contained in the higher harmonics converges
to a definite integral energy as discussed above.
\begin{figure}
\begin{centering}
\includegraphics[width=0.5\columnwidth]{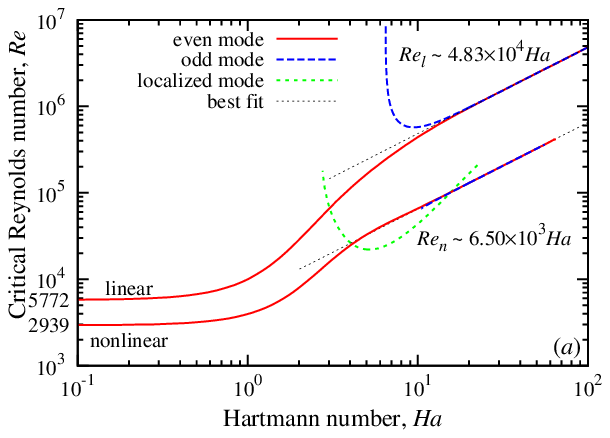}\includegraphics[width=0.5\columnwidth]{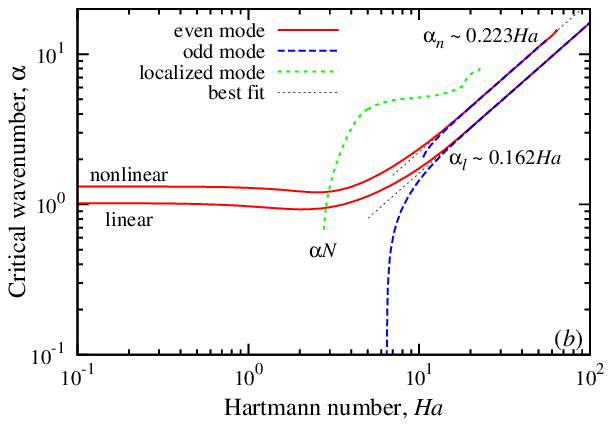}
\par\end{centering}

\begin{centering}
\includegraphics[width=0.5\columnwidth]{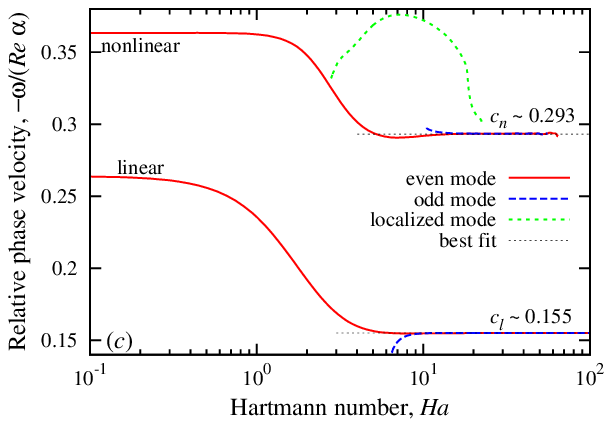}
\par\end{centering}

\protect\caption{\label{fig:crt-Ha} Critical Reynolds number (a), wavenumber (b) and
phase speed (c) for even and odd modes of linear and nonlinear instabilities
against the Hartmann number. The effective wavenumber of the localized
(nonlinear odd) mode shown in figure \ref{fig:str2}(b) is given by
$\alpha N,$ where $N$ is the number harmonics. }
\end{figure}

The critical Reynolds number and wavenumber for the 2D nonlinear stability
threshold are shown in figure \ref{fig:crt-Ha} together with critical
parameters for linear stability versus the Hartmann number \citep{Hagan-Priede-13b}.
At small Hartmann numbers, instability is associated with the even
mode for which 2D travelling waves appear at $\textit{Re}_{n}=2939.$
This is the 2D nonlinear stability threshold of plane Poiseuille flow
shown in table \ref{tab:PPF}. When the Hartmann number exceeds $\textit{Ha}\approx2.8,$
which is about half of the respective value for the linear instability,
an odd equilibrium mode appears with a large Reynolds number and a
small wavenumber. This long-wave odd mode exists only within a limited
range of Hartmann numbers up to $\textit{Ha}\approx20.$ At $\textit{Ha}\approx10$
another odd mode appears with a slightly higher Reynolds number but
much shorter wavelength. At $\textit{Ha}\approx15$ the Reynolds number
of the latter mode becomes smaller than that for the long-wave mode.
The characteristics of this short-wave odd mode are seen to be closely
approaching those of the original even mode. In a sufficiently strong
magnetic field, the critical Reynolds number and wavenumber for both
nonlinear modes increase with the Hartmann number similarly to the
respective threshold parameters of the linear instability \citep{Hagan-Priede-13b}.
Namely, for $\textit{Ha}\gtrsim20$ the best fit yields 
\begin{eqnarray}
\textit{Re}_{n} & \sim & 6.50\times10^{3}\textit{Ha},\label{eq:rec2}\\
\alpha_{n} & \sim & 0.223\textit{Ha},\label{eq:kc2}\\
c_{n} & \sim & 0.293.\label{eq:cc2}
\end{eqnarray}
It is important to notice that the critical Reynolds number above
is almost an order of magnitude lower than that for the linear instability
$\textit{Re}{}_{l}\sim48\,300\textit{Ha}.$ In the mean-field approximation
using only one harmonic, we find $\textit{Re}_{n}\sim12\,300\textit{Ha},$
which is almost a factor of two higher than the accurate result above
and coincides with the result reported by \citet{LifSht80}. 

Besides the threshold parameters above, the critical 2D travelling
wave can also be characterized by its energy amplitude $A_{E}$ and
the flow rate perturbation 
\begin{equation}
\hat{\psi}_{0}(1)=\int_{0}^{1}(\hat{u}_{0}(z)-\bar{u}_{0}(z))\, dz\label{eq:psi0}
\end{equation}
which are plotted in figure \ref{fig:nln2} versus the Hartmann number.
At large $\textit{Ha}$, the latter is seen to approach $\hat{\psi}_{0}(1)\sim-242,$
which means that the critical flow rate perturbation becomes independent
of the magnetic field strength. This is because the flow rate is determined
by the product of characteristic length and velocity scales, of which
the latter varies inversely with the former. Thus, the Hartmann layer
thickness $\delta\sim h/\textit{Ha}$, which defines the characteristic
length scale in strong magnetic field, cancels out in the flow rate
perturbation. The same arguments also explain the scaling of the energy
perturbation (\ref{eq:nrg}) inversely with $\textit{Ha},$ which
leads to $A_{E}\sim0.0317\textit{Ha}^{-1/2}$ for the even mode as
well as for the odd short-wave mode (see figure \ref{fig:nln2}).
The same relation for both instability modes again implies that the
perturbations originating in the Hartmann layers at the opposite walls
do not affect each other in a sufficiently strong magnetic field \citep{Hagan-Priede-13b}. 

Streamlines of the critical finite-amplitude perturbations of both
symmetries are plotted in figure \ref{fig:str2} for $\textit{Ha}=10,20.$
Note that the streamlines of the odd mode are mirror-symmetric with
respect to the mid-plane $z=0$ whereas those of the even mode posses
a central rather than a $z-$reflection symmetry. As discussed in
the description of the numerical method, this is because all stream-function
harmonics of the odd mode are odd functions of $z$ whereas those
of the even mode have alternating $z$ parities. It is interesting
to note that the long-wave odd mode at $\textit{Ha}=10$ represents
a localized disturbance consisting of a pair of mirror-symmetric vortices.
As discussed above, this long-wave equilibrium state disappears at
$\textit{Ha}\gtrsim20.$ The short-wave state, which replaces the
former at higher Hartmann numbers, is seen in figure \ref{fig:str2}
to differ from that of the even mode only by a half-wavelength shift
between the top and bottom parts of the channel.

\subsection{Linear stability of travelling waves to 2D disturbances}

\begin{figure}
\begin{centering}
\includegraphics[width=0.5\textwidth]{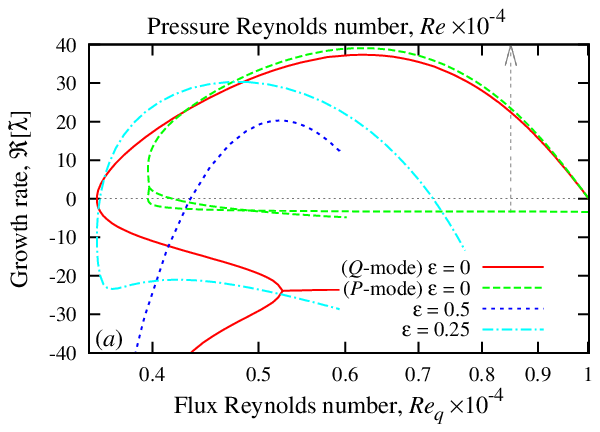}\includegraphics[width=0.5\textwidth]{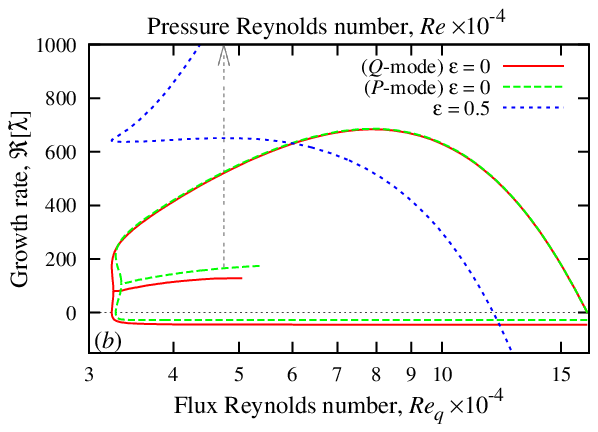}
\par\end{centering}

\protect\caption{\label{fig:grt-Ha}Growth rates of 2D instability modes with various
detuning parameters $\epsilon$ versus Reynolds number for $\textit{Ha}=1$
(a) and $\textit{Ha}=5$ (b). All modes except the fixed-pressure
$(P)$ mode are plotted against the flux Reynolds number $Re_{q}$
(\ref{eq:Req}). The growth rate of the $P$-mode is plotted against
the original Reynolds number based on the mean pressure gradient. }
\end{figure}

Growth rates of of infinitesimal disturbances of the even travelling-wave
mode are plotted in figure \ref{fig:grt-Ha} for various detuning
parameters and $\textit{Ha}=1;5.$ As discussed in Sec. \ref{sub:2D-states},
the subcritical equilibrium states bifurcating from the base flow
are invariably unstable. This is confirmed by the positive growth
rate of the fundamental fixed-flow-rate ($Q$) mode , which is seen
in figure \ref{fig:grt-Ha}(a) to persist down the lowest Reynolds
number $\textit{Re}_{q}$ based on the flow rate. The growth rate
of this mode, which is positive and purely real for the unstable lower
travelling-wave branch turns negative as the solution passes through
the turning point at the critical $\textit{Re}_{q}\approx3557$ and
proceeds to the upper branch. The lower branch can be identified as
the one starting from the linear instability threshold at the largest
$Re$ while the upper branch as that terminating at $Re_{q}\approx6000.$
Further this purely negative eigenvalue merges with another similar
eigenvalue, which results in two complex-conjugate eigenvalues with
negative real part. Note that a purely real growth rate describes
disturbances that travel synchronously with the same phase speed as
the background wave, whereas the complex growth rate describes asynchronous
disturbances which lead to quasi-periodic solutions in the laboratory
frame of reference and periodic solutions in the co-moving frame of
reference. The change of stability at the lowest value of $\textit{Re}_{q}$
corresponds to a saddle-node bifurcation which occurs at fixed flow
rate and gives rise to a stable and unstable branch of travelling-wave
states. Owing to the phase invariance of the travelling-wave solution
there is always a neutrally stable disturbance corresponding to a
small phase shift, i.e. a small shift in $x$ of the original solution.

The growth rate of the fixed-pressure ($P$) mode is plotted in figure
\ref{fig:grt-Ha}(a) against the original Reynolds number based on
the pressure gradient. This mode is also associated with a saddle-node
bifurcation, which in this case occurs at the lowest value $Re.$
As for the $Q$-mode, there is a perturbation with purely real eigenvalue
which changes from positive to negative as the travelling-wave solution
passes through the turning point at $Re\approx3962.$ But in contrast
to the fixed flow rate, this is not the leading but the second largest
eigenvalue. Moreover, conversely to the $Q$-mode, this eigenvalue
is negative on the lower branch and positive on the upper branch.
At the same time, the leading eigenvalue, which is positive on the
lower branch, as for the $Q$-mode, also remains positive after the
turning point when the solution passes to the upper branch. Thus,
when the flow is driven by fixed pressure gradient, there are two
unstable modes with real eigenvalues on the upper travelling-wave
branch just after the turning point. At $Re\approx3972$ these two
unstable modes merge forming a pair of modes with complex-conjugate
eigenvalues. At $Re\approx4170$ the real part of the complex-conjugate
eigenvalues turns negative and the upper branch of the $P$-mode becomes
stable as for the fixed flow rate considered before. These stability
characteristic of $Q$- and $P$-modes for $Ha=1$ are not essentially
different from those of the non-magnetic Poiseuille flow \citep{Pugh-Saffman88,Soibelman-Meiron-91,Casas-Jorba-12}. 

As noted in \S\ref{sub:2Dlstab}, the distinction between fixed-pressure
and fixed-flow-rate cases vanishes for detuned modes $(\epsilon\not=0).$
Growth rates of two such modes with $\epsilon=0.25$ and $\epsilon=0.5$
are plotted in figure \ref{fig:grt-Ha}(a). The former is seen to
destabilize the lower branch of travelling-wave states only at sufficiently
subcritical Reynolds numbers. The latter is a subharmonic disturbance
which does the same for the upper branch. Disturbances with $\epsilon=0.75$
and $\epsilon=1$ are found to be stable $(\Re[\tilde{\lambda}]<0)$
and thus not shown here.

As seen in figure \ref{fig:grt-Ha}(b), the basic features of the
$P$-mode at $\textit{Ha}=5$ remain essentially the same as those
for $\textit{Ha}=1.$ At the same time, the $Q$-mode changes significantly
and closely approaches the $P$-mode. Also, the difference between
the pressure and flux Reynolds numbers diminishes with the increase
of Hartmann number. As discussed in the previous section, this is
due to the localization of perturbations at the walls which results
in the reduction of the flow rate perturbation with the increase of
Hartman number (see figure \ref{fig:nln2}b). Instability of both
solution branches at the turning point implies that additional, possibly
quasi-periodic, equilibrium states may exist and also extend towards
more subcritical Reynolds numbers as speculated by \citet{Barkley-90}
for the non-magnetic case. It is unclear whether such solutions bifurcate
from the upper branch as an excessive number of harmonics is required
to produce reliable results at higher Reynolds numbers and larger
travelling-wave amplitudes.

\section{\label{sec:Sum}Summary and conclusions}

The present study was concerned with 2D nonlinear travelling-wave
states in MHD channel flow subject to a transverse magnetic field.
Such states are thought to mediate transition to turbulence, which
is known to take place in this flow at Reynolds numbers more than
two orders of magnitude below the linear stability threshold. Using
the Newton-Raphson method, we determined the extension of such 2D
nonlinear travelling waves below the linear stability threshold, at
which they appear as the result of a subcritical bifurcation. Starting
from the non-magnetic plane Poiseuille flow, where these travelling-wave
states extend from the linear stability threshold $\textit{Re}{}_{l}=5772$
down to $\textit{Re}{}_{n}=2939,$ and gradually increasing the magnetic
field, we found that in a sufficiently strong magnetic field $(Ha\gtrsim10)$
such states extend down to $\textit{Re}{}_{n}\sim6.50\times10^{3}\textit{Ha.}$
Although this Reynolds number is almost an order of magnitude lower
than the linear stability threshold of the Hartmann flow, it is still
morethan an order of magnitude greater than that at which turbulence
is observed in this type of flow.

Besides the solution which evolves from plane Poiseuille flow as the
magnetic field is increased, we found another inherently magnetohydrodynamic
solution which bifurcates subcritically from the Hartmann flow at
$Ha\approx6.5.$ The two solutions differ by their $z$-symmetries:
the harmonics of the transverse velocity are odd functions of $z$
for the latter and they have alternating parities starting with even
one for the former. The odd-symmetry solution was found to have two
branches. The first branch, which exists in a limited range of the
magnetic field strength with $2.8\lesssim Ha\lesssim20,$ has a relatively
long wavelength and consists of a pair of intense localized vortices.
The second branch, which emerges at $Ha\approx10$ with a much shorter
wavelength, has very similar characteristics to the original hydrodynamic
branch, and becomes practically indistinguishable from the latter
when $Ha\gtrsim20.$ We also showed that the lower-amplitude 2D travelling
waves, as in the non-magnetic case, are invariably unstable with respect
to 2D infinitesimal superharmonic disturbances down to the limiting
value of the Reynolds number based on the flow rate. 

Subcritical 2D travelling waves in the Hartmann flow are also likely
to be unstable to more general three-dimensional disturbances similar
to those considered by \citet{Orszag-Patera-83} in the non-magnetic
case. Three-dimensional equilibrium states bifurcating either from
2D travelling waves \citep{Ehrenstein-Koch-91} or infinity \citep{Waleffe-01},
as in plane Poiseuille flow, may extend to significantly lower Reynolds
numbers and, thus, provide a more adequate threshold for the onset
of turbulence in the Hartmann flow. Such a possibility is supported
by the interaction of only two mirror-symmetric oblique waves with
the resulting 2D second harmonic considered by \citet{ZinSht87},
who found, for the Hartmann layer, $R_{n}\approx4670.$ As for the
2D waves considered in this study, it is likely that a more adequate
three-dimensional model including a sufficient number of higher harmonics
could result in a substantially lower $R_{n}.$ 

\begin{acknowledgements}
J.H. thanks the Mathematics and Control Engineering Department at Coventry University for funding his studentship. J.P. is grateful to Chris Pringle for many useful discussions.
\end{acknowledgements}

\end{document}